\documentclass[%
 aip,
 amsmath,amssymb,
reprint,
]{revtex4-1}

\usepackage{graphicx}
\usepackage{dcolumn}
\usepackage{bm}

\usepackage[utf8]{inputenc}
\usepackage[T1]{fontenc}
\usepackage{mathptmx}
\usepackage[T1]{fontenc} 
\usepackage{amsmath,color}
\usepackage{amssymb}
\usepackage{amsfonts}
\usepackage{amsthm}
\usepackage{mathtools}
\usepackage[title]{appendix}
\usepackage{hyperref}
\usepackage{braket}
\usepackage{xcolor}
\usepackage{etoolbox}
\usepackage{soul}

\newcommand*{\citen}[1]{%
  \begingroup
    \romannumeral-`\x 
    \setcitestyle{numbers}%
    \cite{#1}%
  \endgroup
}

\usepackage{algorithm}
\usepackage{algpseudocode}
\usepackage{dsfont}

\usepackage{bbold}

\begin{document}

\title{Molecular Chirality Quantification: Tools and Benchmarks}

\author{Ethan Abraham}
\email{abrahame@sas.upenn.edu}
\affiliation{Department of Physics and Astronomy, University of Pennsylvania, Philadelphia, Pennsylvania 19104, USA }
\author{Abraham Nitzan}
\affiliation{Department of Chemistry, University of Pennsylvania, Philadelphia, Pennsylvania 19104, USA }
\affiliation{School of Chemistry, Tel Aviv University, Tel Aviv 69978, Israel}

\date{\today}

\begin{abstract}
\vspace{1 cm}
Molecular chirality has traditionally been viewed as a binary property where a molecule is classified as either chiral or achiral, yet in the recent decades mathematical methods for quantifying chirality have been explored. Here we use toy molecular systems to systematically compare the performance of two state of the art chirality measures (1) the Continuous Chirality Measure (CCM) and (2) the Chirality Characteristic ($\chi$). We find that both methods exhibit qualitatively similar behavior when applied to simple molecular systems such as a four-site molecule or the polymer double-helix, but we show that the CCM may be more suitable for evaluating the chirality of arbitrary molecules or abstract structures such as normal vibrational modes. We discuss a range of considerations for applying these methods to molecular systems in general, and we provide links to user-friendly codes for both methods. We aim this paper to serve as a concise resource for scientists attempting to familiarize themselves with these chirality measures or attempting to implement chirality measures in their own work.

\end{abstract}
\maketitle
	
	
\section{Introduction}
\label{intro}

Chirality is defined by non-superimposable mirror images, manifesting in the absence of mirror symmetry.\footnote{Rigorously, chirality is defined by the absence of improper symmetry in general, which includes inversion $\hat{i}$ and $\hat{S}_n$ with even $n$, in addition to mirror symmetry $\hat{\sigma}$.} Since symmetry properties are binary in the sense that an object either possesses or does not possess a particular symmetry element, molecular chirality has historically been a qualitative subject. But with molecular chirality playing a central role in a variety of optical, biological, and chemical phenomena, methods for quantifying chirality could have great utility for research that attempts to treat chirality as a physical variable. Renewed interest in molecular chirality and its quantification has grown with the surge of observations associated with the chiral induced spin selectivity phenomenon.\cite{CISS_overview,CISS_theory}

As an example of the utility of chirality measurement, a recent study by Piaggi \textit{et al.}\cite{critical_chiral} used molecular dynamics simulations of a liquid composed of a simple four-site molecular model to demonstrate the existence of symmetry-breaking phase transitions from a supercritical racemic liquid into subcritical D-rich and L-rich liquids. These phases were observed via the distribution of chirality values of the liquid's molecular constituents, each of which could assume an arbitrary conformation (and thus chirality) at any point throughout the simulation. As such, the success of this study depended on the quality and implementation of a chirality measurement tool. 

Chirality measures have recently been applied in a wide variety of contexts such as the analysis of protein structure in biology,\cite{bio_chir_char, bio_chir_char_previous, chiral_ramachandran} the implication of chiral fermions in topological condensed matter phenomena,\cite{chiral_fermions, chiral_fermions2} and the role of chirality in photoelectron spectroscopy.\cite{chiral_momentum} In our own work, chirality measurement has been used to explore the role of mechanical twist on the thermal conductivity of polymer wires,\cite{twist_paper} and to study the correlation of chiral normal vibrational modes to the chirality of the underlying molecular structure.\cite{chiral_modes_helices}  

Approaches to chirality quantification have fallen into two general categories. The first is based on geometric overlap between a structure and its enantiomer, where high overlap is indicative of low chirality, and larger difference is indicative of higher chirality. This idea was first explored several decades ago when Gilat and Schulman defined a dimensionless chirality measure as one minus the fractional volume overlap of the two enantiomers.\cite{chiral_efficiency} A different and more general approach to the overlap-based chirality measure was developed by Avnir \textit{et al.},\cite{c2,c3} whose Continuous Symmetry Measure (CSM) relies on vector inner products instead of volume overlap.\cite{csm} This leads to the Continuous Chirality Measure (CCM), the first tool addressed in this paper.\cite{c2,c3} 

An alternative approach to chirality quantification is based on a scalar triple product of sequential vectors (see example below), utilizing the fact that an ordered set of three vectors is achiral if and only if they lie in the same plane, in which case the scalar triple product measures zero volume. In all other cases the magnitude of the scalar triple product is nonzero, achieving a maximum when the three vectors are mutually orthogonal. Like the overlap-based method, the idea of using the scalar triple product to measure chirality can be traced back several decades,\cite{early_chir_char} but has been more recently refined by Sidorova \textit{et al.} in their application to biological proteins.\cite{bio_chir_char_previous, bio_chir_char} A strength of this measure is that the anti-commutativity of the scalar triple product means that opposite enantiomers will have opposite signs.\footnote{It has been suggested by mathematicians that there may be no universal measure of handedness and that handedness is just an agreed upon convention for labeling. The handedness assigned by the Chirality Characteristic to the structures in this work is in accord with the conventional classification of helices as either left-handed or right-handed. For certain non-helical structures, or special points along the continuous deformation of a helix into its enantiomer, this classification may be problematic.} This is the basis of the Chirality Characteristic ($\chi$), the second tool addressed in this paper. Though incredibly useful for certain simple systems as well as systems with linear connectivity (e.g. polymers), a weakness of the scalar triple product approach is that it is in general difficult to assign a definite order of sequential vectors, rendering the measure ill-defined. With greater mathematical sophistication, Harris \textit{et al.} presented a more general way of defining the scalar triple product based on multipole moments of the molecule's density function,\cite{multipole_chiral} but here we address simpler choices for the triple-product vectors at the expense of some generality. 

\begin{figure}[tbh]
\includegraphics[width=1\columnwidth]{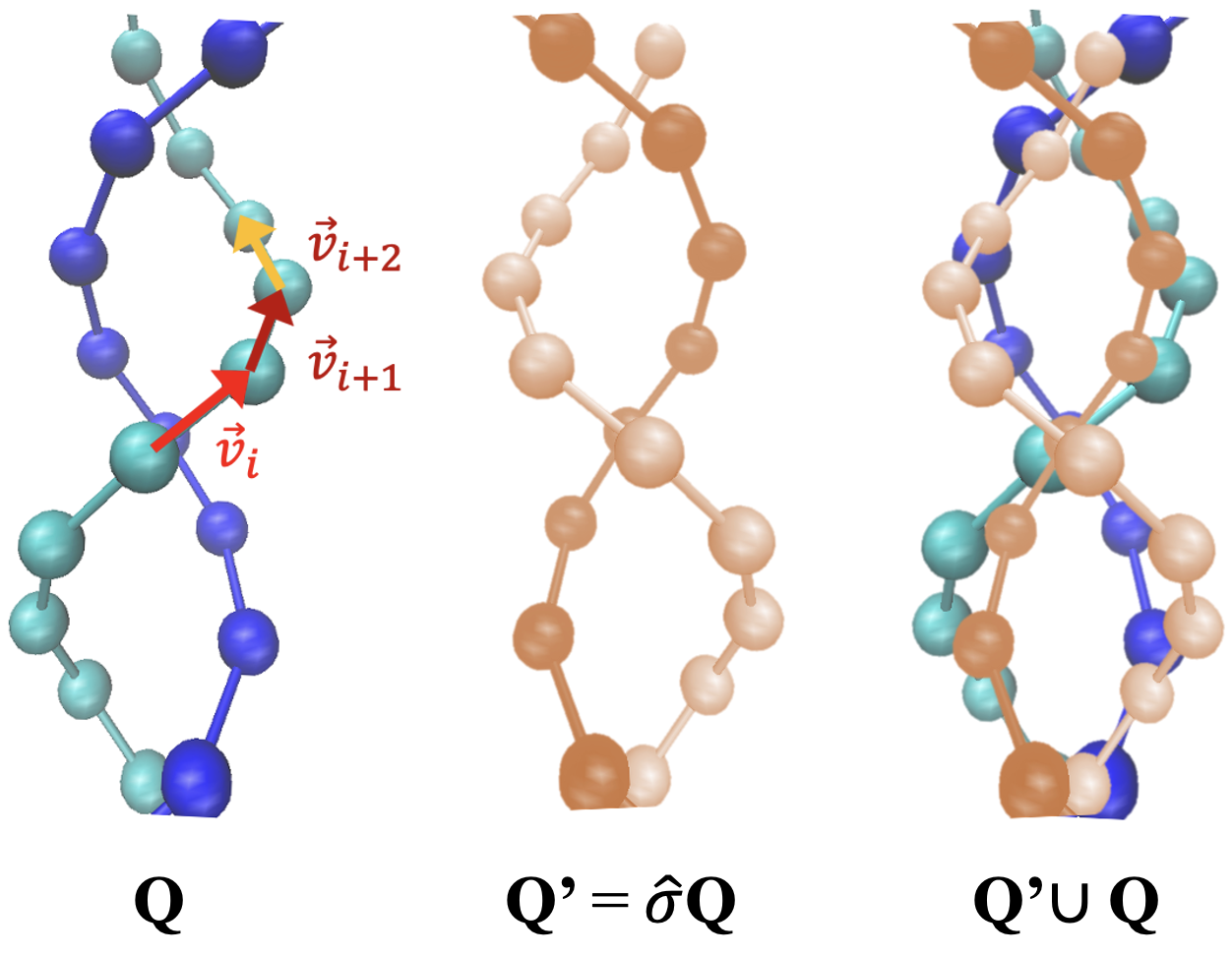}
\caption{Left: ball-and-stick cartoon of a unit turn of a right-handed two polymer double helix, the one of the toy molecular system on which chirality measures are evaluated in this work. Different chains are denoted by different colors. Three sequential separation vectors are highlighted. Middle: Mirror plane reflection of the same structure to generate its left handed enantiomer. Right: overlay of the two enantiomers on a common center of mass origin.}
\label{FIG1}
\end{figure}
\hyperlink{FIG1}{Figure 1} provides intuition for both approaches. The left panel depicts a right-handed double helix, which is one of the toy molecular system on which we will evaluate the CCM and the Chirality Characteristic in this work. It can be seen from the vectors in the left panel that the magnitude of the triple product $(\textbf{v}_{i}\times\textbf{v}_{i+1})\cdot\textbf{v}_{i+2}$ reports the helicity of the structure,\footnote{The notion of helicity as defined here is analogous with the conventional definition of helicity in physics as the projection of a pseudovector on a vector, such as the projection of angular momentum onto linear momentum.} while the sign of this quantity distinguishes between the enantiomers. This is the essence of the scalar-triple-product-based chirality measures such as the Chirality Characteristic. The middle panel shows the left-handed enantiomer of the same helix, formed by a mirror reflection. The right panel shows an overlay of both mirror images on a common center of mass origin. For an achiral structure, the second structure could be rotated such that the overlap is perfect. For a more chiral structures, the overlap will be less perfect. This is the essence of the overlap-based chirality measures such as the CCM. 

In this paper, we evaluate these chirality measures on toy molecular systems in order to compare their performance. In \hyperref[details]{Sec. II} we present the mathematical formulas for these measures. In \hyperref[toy]{Sec. III}, we discuss the application of such measures using two simple molecular models. The first model is a linear, four-site molecule, and the second is a polymer double helix. Our findings are summarized in \hyperref[conclusion]{Sec. IV}. 

\section{Chirality Measures}
\label{details}

Here we will provide two state state of the art tools to measure the chirality of molecular systems. The two tools presented correspond to the two general classes of chirality measures discussed above, with the first based on the overlap of enantiomers and the second based on the scalar triple product.

{\it Continuous Chirality Measure.} Let $\textbf{Q}=\{\textbf{q}_1,...,\textbf{q}_n\}$ denote the $n$ atomic coordinates of the molecule of interest relative the center of mass. Then the CCM is given by
\begin{equation}\tag{1}\label{ccm_mol} CCM = 1 - \frac{\max_{\hat{\sigma},\hat{P}}\sum_{i = 1}^{n} \textbf{q}_i\cdot(\frac{\hat{\mathds{1}}+\hat{\sigma}}{2})\hat{P}\textbf{q}_i}{\sum_{i = 1}^{n} \lVert \textbf{q}_i \rVert^2},\end{equation} 
where $\hat{\mathds{1}}$ is the identity operator, $\hat{\sigma}$ is a reflection operator, and $\hat{P}$ is a permutation operator ($\hat{P}\textbf{q}_i=\textbf{q}_j).$\cite{c2} For derivation and further discussion of the CCM, see Ref. \citen{c2} and \citen{c3}. Naively, the intuition for the form of Eq. \ref{ccm_mol} is that the numerator of the second term reports the maximal inner product between the original structure $\textbf{Q}$ and the \textit{nearest achiral structure} $\frac{1}{2}(\hat{\mathds{1}}+\hat{\sigma})\hat{P}\textbf{Q}$ formed by averaging the original molecule $\textbf{Q}$ with its enantiomer $\textbf{Q'}=\hat{\sigma}\hat{P}\textbf{Q}$.\footnote{As noted in Ref. Ref. 14, other improper symmetry measures such as $\hat{i}$ and $\hat{S}_n$ (with $n$ even) should be considered in addition to $\hat{\sigma}.$ However, it has been found that in the vast majority of cases (>$90\%$) the operation $\hat{\sigma}$ yields the correct (minimum) CCM. In this work, we consider only $\sigma$ for simplicity.} Although not enforced in all past implementations of the CCM, we require that $\hat{P}\textbf{Q}=\textbf{Q}$, meaning that $\hat{P}$ must be restricted to permutations that are invariant with respect to groups of chemically equivalent atom. Thus, with the denominator serving normalization purposes, the second term in Eq. \ref{ccm_mol} reports a dimensionless measure of the similarity between $\textbf{Q}$ and its nearest achiral structure.  The CCM thus returns 0 for an a achiral molecules and at most 1 for chiral molecules. For an analytical discussion of how to maximize the second term over the $\hat{\sigma},$ see Ref. \citen{csm}. Maximization over the $\hat{P}$ is performed computationally. For such CCM calculations, we use the \texttt{cosymlib} open source software released by developers of the CCM to which a link is provided in the data availability section below.

{\it Chirality Characteristic.} Let $\textbf{Q}=(\textbf{q}_1,...,\textbf{q}_n)$ denote an ordered set of $n$ atomic coordinates of the molecule of interest relative the center of mass. Let $\textbf{v}_i = \textbf{q}_{i+1}-\textbf{q}_{i}$ for $i=1,...,n-1.$ Then we define the Chirality Characteristic as
\begin{equation}\tag{2}\label{chi_mol} \chi =\frac{\sum_{i = 1}^{n-3} (\textbf{v}_i\times\textbf{v}_{i+1})\cdot\textbf{v}_{i+2}}{\sum_{i = 1}^{n-3} \Vert\textbf{v}_i\rVert\lVert\textbf{v}_{i+1}\rVert\lVert\textbf{v}_{i+2}\rVert}.\end{equation}
This expression reports a normalized average of scalar triple products of successive bond vectors throughout the molecule. Note that unlike for the CCM, where the labeling of the $\textbf{q}_i$ could be arbitrary, here we see that $\chi$ is dependent on the ordering of the $\textbf{q}_i$. Thus we see that $\chi$ is natural to evaluate in molecules with linear connectivity but difficult to define on molecules with branched structures.  Note that in Ref. \citen{bio_chir_char}, a different normalization method was used; we instead follow Ref. \citen{critical_chiral} in using the normalizing denominator that ensures that the Chirality Characteristic is dimensionless (ranging from $-1$ to 1). We note that unlike the CCM which reports the same value for either enantiomer, the $\chi$ assigns opposite sign to opposite enantiomers. In particular, right-handed helices have positive $\chi$ values and left handed helices have negative $\chi$ values. We provide link to an R script used to compute this quantity in the data availability below.  

\section{Performance on Toy Molecular Systems}
\label{toy}

In this section we show results for evaluating these chirality measures on two toy molecular systems. The first such system is the four-site molecular model, and the second is the polyethelene double helix. Finally, we comment briefly on how the measures perform on a single helix, as compared to a double helix.

\begin{figure}[tbh]
\includegraphics[width=1\columnwidth]{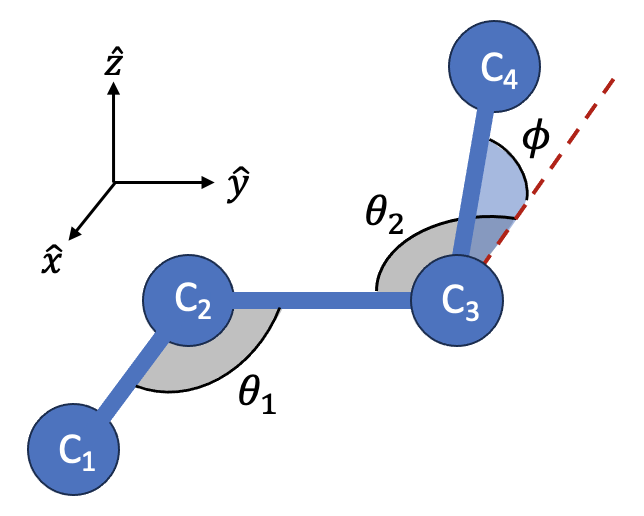}\caption{Schematic of a linearly connected four-carbon system. Holding bond-lengths constant, all conformations can be specified by three parameters $(\theta_1,\theta_2,\phi)$ where C$_2$ and C$_3$ are fixed, and angle $\angle$C$_1$C$_2$C$_3$ is confined to the $xy$-plane. We assume that all three bonds have length $l$.}
\label{FIG2}
\end{figure}

{\it Four Carbon System.} \hyperref[FIG2]{Figure 2} shows the simple four-site atomic model that we use to evaluate the aforementioned chirality measures. The atoms in \hyperref[FIG2]{Fig. 2} are labeled as Carbons, but the identity of the atoms in the model do not affect either chirality measurement provided that the atoms are identical.\footnote{If the atoms are not identical, then one must confront the question of how such differences should be accounted for, such as weighting by mass.} Assuming all three bonds have length $l$, we note that the conformation of such a structure is fully determined by three parameters. We orient our coordinate system so that the origin is at the location of C$_2$, C$_3$ lies on the $y$-axis, and C$_1$ lies in the $xy$-plane. Then as indicated in \hyperref[FIG2]{Fig. 2}, any conformation is specified by three angles: $\theta_1=\angle$C$_1$C$_2$C$_3$, $\theta_2-$the angle made by the C$_3$-C$_2$ bond with the projection of the C$_3$-C$_4$ bond on the $xy$-plane, and $\phi$ which is the the angle between the C$_3$-C$_4$ bond and the $xy$-plane.

We note that for this simple system, $\chi$ has has a concise analytical treatment $\chi^{4\text{-site}}=l^{-3}(\textbf{r}_{12}\times\textbf{r}_{23})\cdot\textbf{r}_{34}=-$sin$\theta_1$sin$\phi$, while the CCM lends itself to no such concise interpretation. 

\hyperlink{FIG3}{Figure 3} shows the results for evaluating both the CCM and $\chi$ for various choices $(\theta_1,\theta_2,\phi).$ In the top panels $\theta_1$ and $\theta_2$ are fixed while $\phi$ varies, while in the bottom panels $\theta_2$ and $\phi$ are fixed while $\theta_1$ is allowed to vary. Note that structures examined based on the angles chosen return negative $\chi$ values. We are therefore required to compare the CCM to $-\chi$ (or $|\chi|$) in order to make such a comparison meaningful. Since the CCM reports positive values for both enantiomers, sign considerations are moot.\\Apart from the quantitative results, which can serve as benchmark values for future chirality calculations, we note several qualitative observations: 

\begin{figure}[tbh]
\includegraphics[width=1\columnwidth]{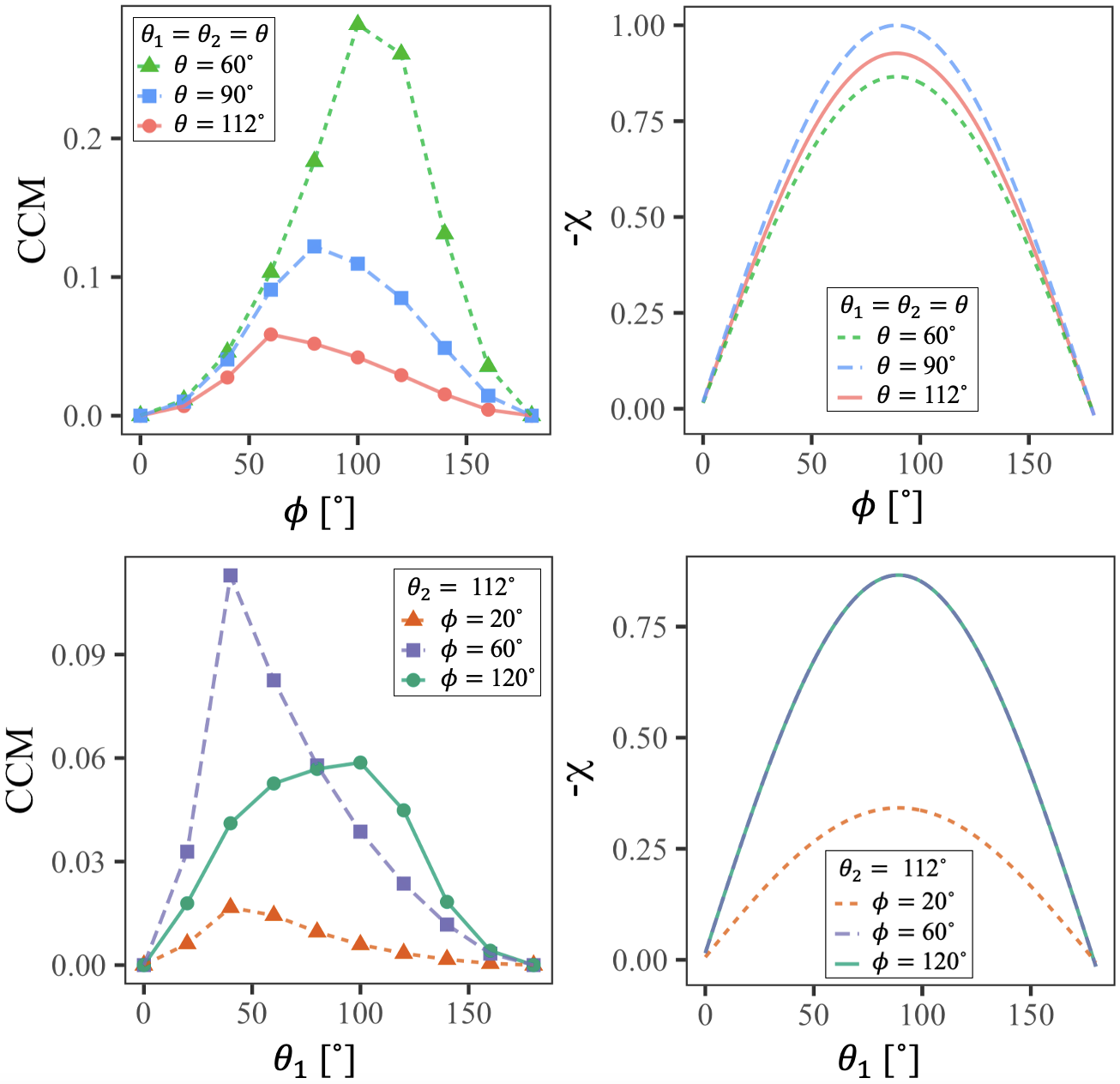}
\caption{Evaluation of the Continuous Chirality measure (CCM) and Chirality Characteristic ($\chi$) on the four-carbon system for selected values $(\theta_1,\theta_2,\phi)$. The left panels show the CCM results while the right panels show the results for (negative) $\chi$. In the top panels, $\phi$ varies as the independent variable while the values $\theta_1=\theta_2=\theta$ are held fixed at $60^{\circ}$ (green), $90^{\circ}$ (blue), and $112^{\circ}$ (red). For the bottom panels, $\theta_1$ is allowed to vary as the independent variable while $\theta_2=112^{\circ}$ is fixed, and also $\phi$ is held fixed at $20^{\circ}$ (orange), $60^{\circ}$ (indigo), and $120^{\circ}$ (green).}
\label{FIG3}
\end{figure}
\begin{quote}
(a) If $\phi$ or $\theta_1$ are equal to $0^{\circ}$ or $180^{\circ}$, the structure is planar and therefore achiral. Both measures correctly return zero chirality for such cases, as indicated by the endpoints of each line in \hyperlink{FIG3}{Fig. 3}.\\
(b) As the independent angle varies from $0^{\circ}$ to $180^{\circ},$ in all cases the chirality magnitude reaches one local and global maximum. However, while for $\chi$ this maximum occurs exactly halfway through the angle's domain (at $90^{\circ}$), for the CCM the maximum occurs at different points depending on the other two angles.\\
(c) For $\chi$, the lines for $\chi=60^{\circ}$ and $\chi=120^{\circ}$ coincide perfectly because the sines of both angles are equal. This is not the case for the CCM.  
\end{quote}

Finally, we note the importance of restricting the permutation operator $\hat{P}$ to permutations of chemically equivalent elements as mentioned in \hyperlink{details}{Sec. II.} In this system, C$_1$ and C$_4$ are chemically equivalent and so are C$_2$ and C$_3$. Note that although all atoms are identical, the atoms C$_1$ an C$_3$ are not chemically equivalent. If the permutation operator permuted C$_1$ with C$_3$, it can be seen that even if $\textbf{Q}$ is achiral, $\hat{\sigma}\hat{P}\textbf{Q}$ may not be an enantiomer of $\textbf{Q},$ and the CCM would no longer be as chemically meaningful.

\begin{figure}[tbh]
\includegraphics[width=1\columnwidth]{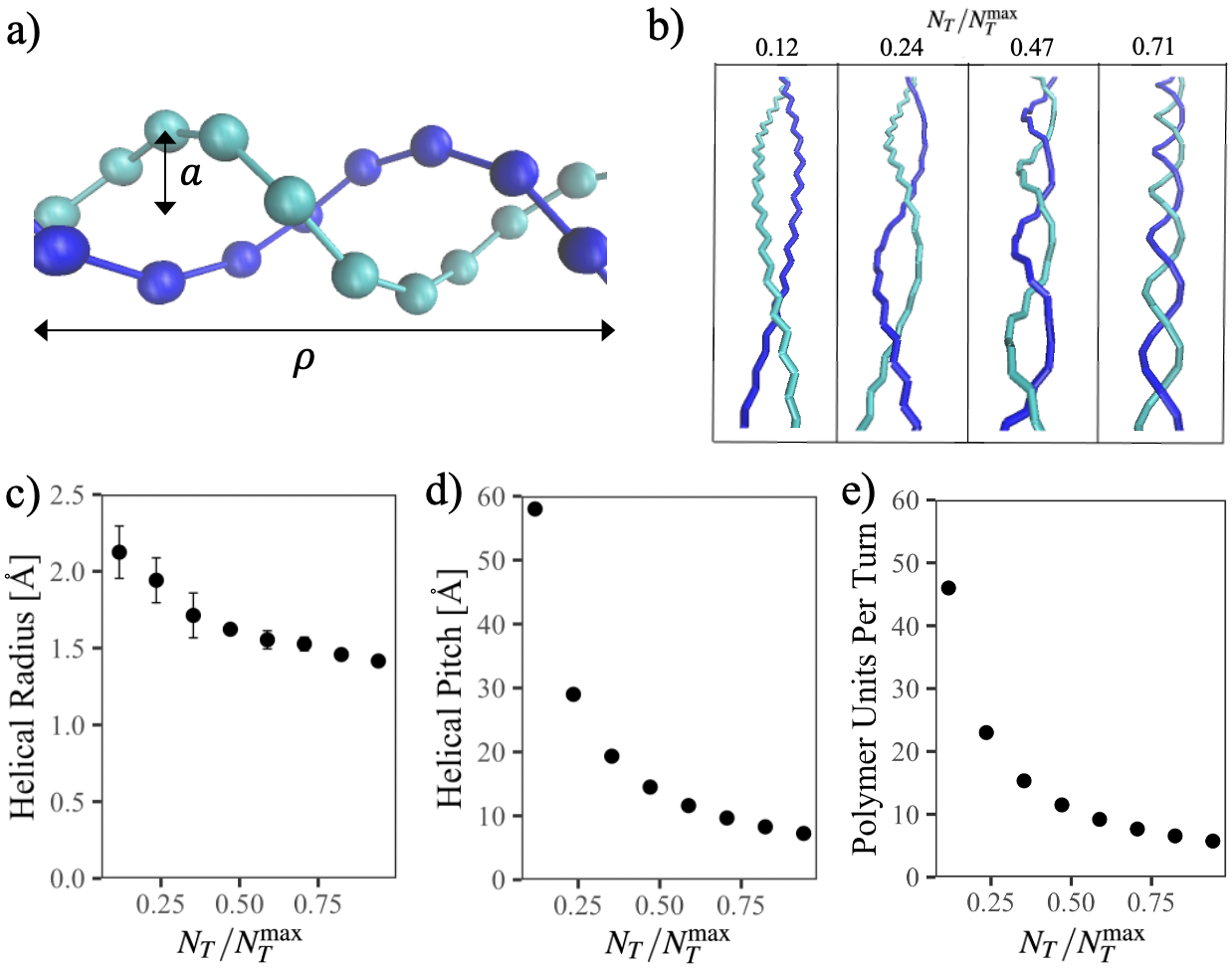}
\caption{(a) Ball-and-stick cartoon of a unit turn of a right-handed polyethelene (PE) double helix, where different chains are denoted by different colors. The double-helix is charecterized by the radius $a$ and the pitch $\rho$ as well as the number of units per helical turn $n$. (b) Snapshots of a model double-helical polyethylene (PE) wire containing various levels of twist. (c,d,e) Estimated helical parameters (radius $a$, pitch $\rho$, polymer units per turn $n$) as a function of twist for the PE model. Note that the helical turn has been sampled from a polymer held at its natural untwisted length as defined in Ref. 11. Twist is reported as the fraction of maximal twist $N_T/N_T^{\text{max}}$, also defined in Ref. 11.}
\label{FIG4}
\end{figure}

{\it Polymer Double Helix.} The second system on which we evaluate the CCM and $\chi$ is the polymer double helix. \hyperref[FIG4]{Figure 4a} shows a unit helical turn of such a helix, which is the structure structure on which we evaluate the chirality measures here. Such a helix can be characterized by the helix radius $a$ and the pitch $\rho$, as well as the number of units per turn of the helix $n$. 

We take a model polyethylene (PE) molecule used in prior work (see Ref. \citen{twist_paper}) in which we performed molecular dynamics (MD) simulations to generate PE double helices with various levels of twist. A United Atom model is used which coarse-grains each CH$_2$ monomer into a single interaction site. As discussed in Ref. \citen{twist_paper}, the level of twist can be characterized by the fraction of the maximal twist that can be attained before bonds begin to break, which we denote $N_T/N_T^{\text{max}}$. \hyperref[FIG4]{Figure 4b} shows snapshots of such a simulated double-helix for various such levels of twist, when the polymer is held at its natural untwisted length as described in Ref. \citen{twist_paper}. We observed that the helical radius $a$ decreases with the level of twist, as shown in \hyperref[FIG4]{Fig. 4c}. The helical pitch $\rho$ and the number of polymer units per helical turn also decrease with twist as expected (see \hyperref[FIG4]{Fig. 4d-e}). However, we note from inspecting \hyperref[FIG4]{Fig. 4b} that the polyethelene model does not form a perfect helix. Specifically, for the lower levels of twist, zig-zag substructures are observed that which deviate from an ideal helical curve (see also Ref. \citen{twist_paper}). We therefore compare our results to that of an ideal helix fitted to the helical parameters observed in the PE model. The position coordinates for the atoms relative to the center of mass for such an ideal single helix are given by \begin{equation}\tag{3}\label{ideal_helix} \textbf{q}_k=\left(a\text{cos}\left(\frac{2\pi k}{n}\right),a\text{sin}\left(\frac{2\pi k}{n}\right),\frac{\rho k}{n}\right)^{T}\end{equation} for $k=0,...,n-1$, and a double helix can easily be formed by adding another such single-helix rotated by $\pi$ about the $z$-axis.

\begin{figure}[tbh]
\includegraphics[width=1\columnwidth]{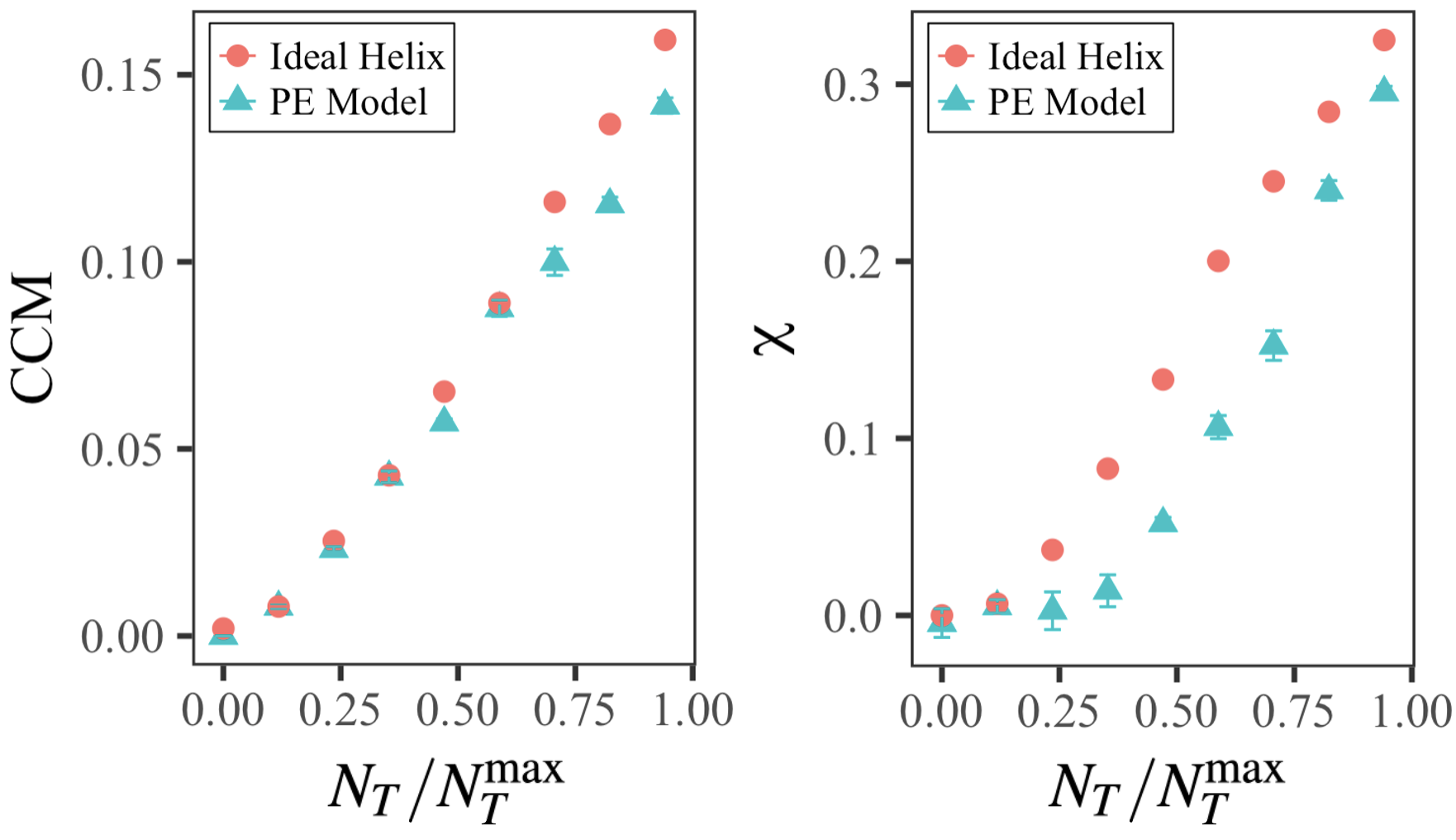}
\caption{The CCM (left panel) and $\chi$ value (right panel) of a unit helical turn for a polymer right-handed double-helix as a function of twist, reported as the PE model's fraction of maximal twist before bonds begin to break $N_T/N_T^{\text{max}}$. Results for the PE model double helix (blue triangles) are compared to results for a corresponding ideal double helix (red circles), where the latter is computed by fitting the mathematical equation of a helix to the parameters shown in \hyperref[FIG4]{Fig. 4}. For the PE model, statistical uncertainties (between arbitrary samplings from simulations) are shown when larger than the symbol sizes. For the idealized helix, such statistics are not relevant.}
\label{FIG5}
\end{figure}

\hyperref[FIG5]{Figure 5} shows results for evaluating the CCM and $\chi$ on both the PE model and analogous ideal helix as a function of the level of twist $N_T/N_T^{\text{max}}$ (where $N_T^{\text{max}}$ from the PE model is used also for the ideal helix). The qualitative similarity between the behavior of the two measures is striking. An important observation is that while the CCM behaves similarly on the ideal and molecular models, $\chi$ deviates at moderate levels of twist. We attribute the reduced $\chi$ value in the PE model to the zig-zag substructure, which diminishes as maximal twist is approached.\cite{twist_paper}

It is interesting to inquire regarding what the plots in \hyperref[FIG5]{Fig. 5} would look like if analogous single helices were used instead of double helices. It can be easily be guessed that the $\chi$ values would be unaffected because the normalization in Eq. \ref{chi_mol} would counteract the decreased number of terms. The effect on the CCM is more difficult to predict. We find that for the ideal helix the plot is qualitatively similar but with magnitudes of roughly 1/3 the double-helical counterpart (results not shown). Similarly, if the measures were evaluated on sections of the double helix of lengths different than $\rho$, the $\chi$ value would be unaffected, but the CCM would vary as discussed in Ref. \citen{chiral_modes_helices}.

\begin{figure}[tbh]
\includegraphics[width=1\columnwidth]{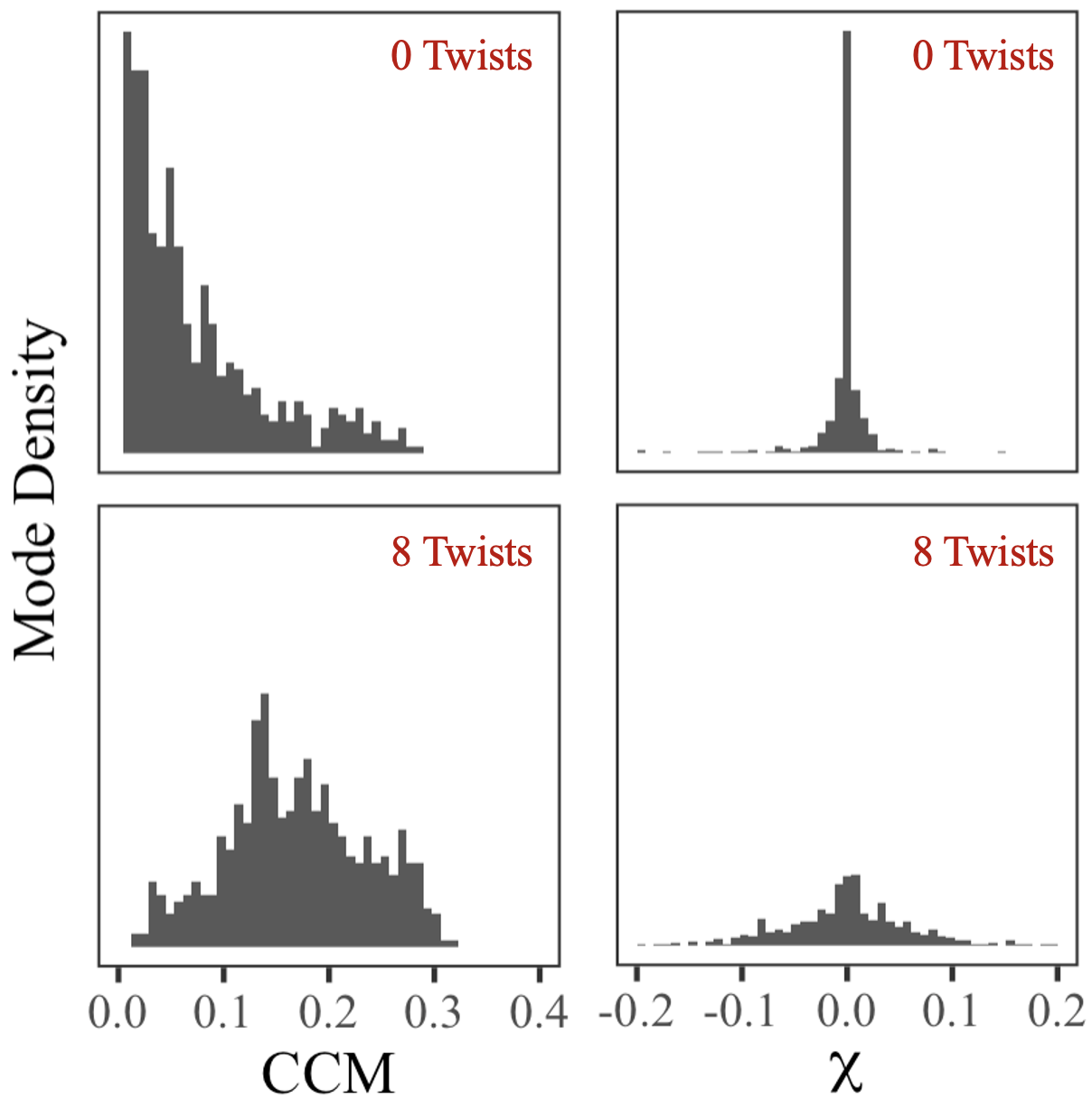}
\caption{Density of normal modes binned by normal-mode CCM value (left panels) and by normal-mode $\chi$ value (right panels) for length $N=98$ left-handed double-helical polyethylene wires that are (top) untwisted and (bottom) twisted to 8 twists (corresponding to $N_T/N_T^{\text{max}}=0.47$). Note that the area under the curve for all four plots are equal, but the $y$-dimension of the right panels are scaled differently compered to the left panels in order to accommodate the sharp peak at $\chi=0$ for 0 twist.}
\label{FIG6}
\end{figure}

\section{Performance on Abstract Structures: Normal Modes}\label{Modes}

It is also worthwhile to compare the performance of both chirality measures on more abstract structures such as molecular normal vibrational modes. Indeed, recent suggestions that chiral phonons may be implicated in the chiral induced spin selectivity (CISS) effects\cite{Spin-Seedback} have focused interest on the connection between the chirality of molecular vibrational modes and that of the underlying molecular structure. We computed the normal vibrational modes for PE double helices using the model described above, using chains of length $N=98$ where $N_T^{\text{max}}\approx17$, as discussed in Ref. \citen{twist_paper}. We use $\textbf{c}_{k,i}$ to denote the mass weighted displacement vector of the $i$th atom in normal mode $k$. On such structures, the definitions of the CCM and $\chi$ have not been well-established, and some creativity is required in order to choose the most natural definition. We follow Ref. \citen{chiral_modes_helices} in defining the CCM of normal mode $k$ by taking $\textbf{q}_i=\textbf{c}_{k,i}$ in Eq. \ref{ccm_mol} and enforcing $\hat{P}=\hat{\mathds{1}}.$ To our knowledge, defining $\chi$ on normal modes has never been done before, and we find that the simplest approach is to take $\textbf{v}_i=\textbf{c}_{k,i}$ in Eq. \ref{chi_mol}

\hyperref[FIG6]{Figure 6} shows the results of such calculation, binning the normal mode spectrum of the model PE (left-handed) double helix by each chirality measure. By comparing the top and bottom panels, we see how twist (i.e. chirality of the underlying molecular structure\footnote{Note that the untwisted wire may not represent a perfectly achiral conformation. This is because for a long and flexible wire, the conformations corresponding to local energy minima can be slightly asymmetric.}) affects the chirality of the normal mode spectrum as measured by the CCM (left) and $\chi$ (right). Note that the plots have different $x$-axis domains because $\chi$ reports negative values for left-handed enantiomers. There are two observations we find noteworthy:
\begin{quote}
(a) In the twisted structure compared to the untwisted structure, both the CCM and $\chi$, report that modes tend to have higher chirality magnitudes.\\
(b) For the CCM, the peak of the distribution shifts with twist, while for $\chi$, the peak remains at zero while the distribution broadens.
\end{quote}

Overall, the data in \hyperref[FIG6]{Fig. 6} indicates that chirality of molecular normal modes is correlated with the chirality of the underlying molecular structures. It may be surprising that that the distribution for $\chi$, which classifies modes by helical handedness, remains centered at zero for the twisted structure despite the left-handed asymmetry of the underlying structure. We have found upon further analysis that the handedness of the helix does introduce asymmetries in the normal mode spectrum, which is the subject of Ref. \citen{chiral_modes_helices}.

\section{Conclusion}
\label{conclusion}

We have discussed two measures of molecular chirality, the Continuous Chirality Measure and the Chirality Characteristic, and their realization when applied to simple model molecular systems. We find that the behavior of both measures is qualitatively similar as structural parameters of the molecules are varied. We conclude by listing technical considerations for choosing between the two measures and applying these measures to molecular systems: 
\begin{quote}
(a) The Chirality Characteristic is only well defined on structures with linear connectivity. On branched structures, the CCM might be a better choice. Further consideration of bond ordering is needed to apply $\chi$ to such structures.\\ 
(b) The CCM returns the same value for either enantiomer while $\chi$ assigns opposite sign to opposite enantiomers. For this reason, although the CCM may be expected to correlate with the magnitude of a chirality-dependent physical effect, it will not predict the direction of the effect.\\
(c) The CCM may be more effective than $\chi$ in measuring the chirality of abstract quantities such as normal vibrational modes. \\
(d) When applying the CCM, care should be taken that the permutation operator $\hat{P}$ is restricted such that the original structure can be accurately compared to its enantiomer without mixing atoms that are not chemically equivalent.
\end{quote}

Though it has already been suggested that the CCM can correlate with the temperature dependence of optical rotation,\cite{chir_opt_rot} as well as transition state enthalpy,\cite{chir_trans_enthalpy} and enantioselective catalysis,\cite{chir_enantio} there is an ongoing effort to explore the wide variety of contexts to which the chirality measures addressed in this work can be applied, as well as the physical meanings of these measures. We hope that this work serves as a useful guide for future efforts to measure the chirality of molecular systems.

\section*{Acknowledgements}
This paper was written in honor of Louis Brus, a creative investigator, a scientific colleague, and a friend of A.N. The research of A.N. is supported by the Air Force Office of Scientific Research under award number FA9550-23-1-0368 and the University of Pennsylvania. E.A. acknowledges the support of the the University of Pennsylvania (GfFMUR Grant). We thank David Avnir for introducing us to his work on the CCM. The authors are grateful to Pere Alemany for useful discussions, and Mohammadhasan Dinpajooh and Claudia Climent for technical assistance and support.

\section*{Data Availability}
The packages used in this work to compute the CCM were released by Pere Alemany \textit{et al.} and are available at:  \url{https://cosymlib.readthedocs.io/en/pere_tutorial/}. An updated version of this code is available at \url{https://zenodo.org/records/4925767}. Another updated resource that includes a CCM calculator and resources to learn about the CCM has been released by Avnir \textit{et al.} and is available at: \url{https://csm.ouproj.org.il/}. Our R script to compute $\chi$  can be found at: \url{https://github.com/eabes23/chi_char/}. The data to generate the twisted polyethylene double helices can be found at \url{https://github.com/eabes23/polymer_twist/}, in accordance with Ref. \citen{twist_paper}.

\section*{References}

\bibliographystyle{aipnum4-1}

\end{document}